\documentstyle[multicol,graphicx,prb,aps]{revtex}
\begin{document}
\draft
\title{
The first-order phase transition between 
dimerized-antiferromagnetic and uniform-antiferromagnetic phases 
in Cu${}_{1-x}$$M_x$GeO${}_3$
}
\author{
T. Masuda,\cite{Masuda-email,AM} 
 I. Tsukada,\cite{IT} and K. Uchinokura\cite{AM}
}
\address{
Department of Applied Physics, The University of Tokyo, 
6th Engineering Bld., 7-3-1 Hongo, Bunkyo-ku, Tokyo 113-8656, Japan 
}
\author{
Y. J. Wang, V. Kiryukhin, and R. J. Birgeneau
}
\address
{
Department of Physics, Massachusetts Institute of Technology, 
Cambridge, Massachusetts 02139
}
\date{\today}
\begin{multicols}{2}[
\maketitle
\begin{abstract}
We have performed detailed magnetic susceptibility measurements
as well as synchrotron x-ray diffraction studies to
determine the temperature vs concentration ($T$ - $x$) 
phase diagram of Cu${}_{1-x}$Mg${}_x$GeO${}_3$. 
We observe clear double peaks in the magnetic susceptibility implying two
antiferromagnetic (AF) transition temperatures 
in samples with Mg concentrations in the range 0.0237 $\le x \le$ 0.0271.    
We also observe a drastic change in the inverse correlation length  
in this concentration range 
by x-ray diffraction. 
The drastic change of the AF transition temperature 
as well as the disappearance of the spin-Peierls (SP) 
phase have been clarified; 
these results are consistent with a first-order phase transition 
between dimerized AF (D-AF) and uniform AF (U-AF) 
phases as reported by T. Masuda {\it et al.} 
\lbrack Phys. Rev. Lett. {\bf 80}, 4566 (1998)\rbrack. 
The $T$ - $x$ phase diagram of Cu${}_{1-x}$Zn${}_x$GeO${}_3$ 
is similar to that of Cu${}_{1-x}$Mg${}_x$GeO${}_3$, 
which suggests that the present phase transition is 
universal for Cu${}_{1-x}M_{x}$GeO${}_3$. 
\end{abstract}
]
\narrowtext

\section{Introduction}
The discovery of the inorganic spin-Peierls (SP) cuprate CuGeO${}_3$
(Ref.~\onlinecite{hase1}) 
has made 
it possible to study systematically the effect of impurities on SP systems. 
The effect of substitution of Zn$^{2+}$ ($S = 0$) for Cu${}^{2+}$ was studied by 
Hase {\it et al.}~\cite{hase2} and the appearance 
of an antiferromagnetic (AF) phase at temperatures 
below the SP transition temperature ($T_{SP}$) was 
reported.~\cite{oseroff,hase3} 
Both dimerization superlattice and AF magnetic peaks
were observed by neutron diffraction measurements 
below the AF transition temperature 
($T_N$) (Ref.~\onlinecite{regnault,sasago,martin}); the coexistence of 
these two seemingly exclusive order parameters was explained 
theoretically by using a phase Hamiltonian method.~\cite{fukuyama} 
Recently, some of the present authors 
studied the transition temperature vs impurity 
concentration ($T$ - $x$) 
phase diagram in Mg${}^{2+}$($S$ = 0)-doped 
CuGeO${}_3$ by means of dc susceptibility 
measurements.~\cite{masuda} 
These authors observed the disappearance of the cusp due to the 
SP transition and the sudden increase of the AF transition temperature 
at an impurity concentration, $x = x_c \sim 0.023$. 
They, therefore, concluded that there was a first-order phase 
transition between dimerized 
antiferromagnetic (D-AF) and uniform antiferromagnetic (U-AF) phases. 
The disappearance of the 
long-range order (LRO) of the dimerization above a critical 
impurity concentration was tentatively explained theoretically 
as a second-order phase transition at $T$ = 0 K.\cite{saito} 
In this paper we report detailed studies on 
the temperature dependence of the magnetic 
susceptibility and, consequently, the 
$T$ - $x$ phase diagram of Cu${}_{1-x}$$M_x$GeO${}_3$ ($M$ = Mg and Zn) 
near $x \sim$ 0.024 (Mg) and 0.020 (Zn).
We have obtained clear evidence for a first-order phase 
transition 
between U-AF and D-AF phases in these nonmagnetic impurity-doped 
systems thereby strengthening the 
conclusion of the previous work.~\cite{masuda} 
We have also used high resolution synchrotron x-ray diffraction techniques 
to clarify the SP phase 
region in the phase diagram. 
As a result we have confirmed the disappearance of the 
SP-LRO at $x \gtrsim x_c$. 
More importantly, we have found that the SP correlation length for a 
given $x$ becomes 
long-range at a lower temperature~\cite{yujie} ($T_{SP}'$) than 
the SP transition temperature previously reported 
and $T_{SP}'$ 
crosses the antiferromagnetic phase boundary at $x \sim 0.024$; 
this gives a phenomenological explanation for the origin of 
the putative first-order phase transition. 


\section{Experimental Details}

\begin{figure}
\begin{center}
\includegraphics*[width=12cm]{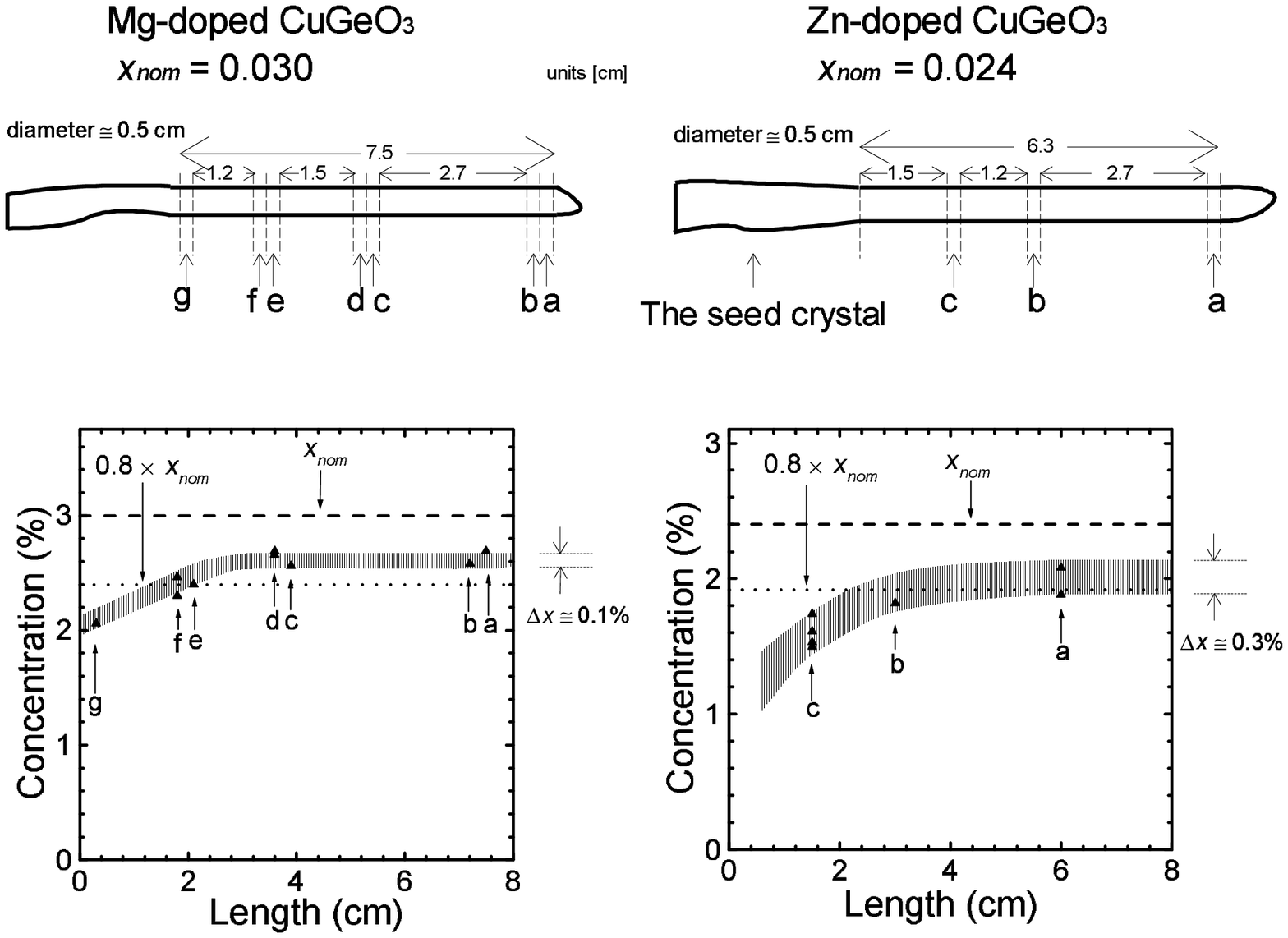}
\caption{
Sketches of bulk single crystals of Cu${}_{1-x}M_x$GeO${}_3$ 
($M$ = Mg and Zn) 
grown
by the FZ method. 
In some cases we analyzed more than one piece. 
The figures in the bottom show the concentration determined 
by ICP-AES vs crystal-growth-length. 
}
\label{fig0}
\end{center}
\end{figure}

All the samples were high quality single crystals 
grown by the floating-zone method; the concentration 
of the Mg$^{2+}$ or Zn$^{2+}$ dilutant 
$x$ was determined by inductively coupled plasma 
atomic emission spectroscopy (ICP-AES). 
To determine the phase boundary we paid special attention to 
any possible inhomogeneity of the impurity concentration. 
We show a rough sketch of typical bulk single crystals of 
Cu${}_{1-x}M_x$GeO${}_3$ ($M$ = Mg and Zn) in Fig.~\ref{fig0}. 
The crystals were grown by using small pure or slightly doped CuGeO${}_3$ 
single crystals 
as the seed crystals. 
When the seed crystal contains a smaller concentration than that of 
starting polycrystalline material, 
the actual grown crystal rod has a concentration gradient at the 
end corresponding to the initial growth stage and then 
the concentration saturate at the later growth stage. 
For the measurements of physical properties we used parts of the 
crystals in the saturated region, so that 
the concentration distribution along the $c$ direction of 
the samples we have studied is 
within the inhomogeneity along the radial direction. 
We estimated the accuracy of the impurity concentration 
from the fluctuation of the saturated concentration 
in a few of the rods. 
From this we concluded that any errors in the concentration are 
within 0.1\% in Cu${}_{1-x}$Mg${}_x$GeO${}_3$ 
and 0.3\% in Cu${}_{1-x}$Zn${}_x$GeO${}_3$ 
in the region of 0.02 $< x <$ 0.03. 
It is apparent that Mg-doped CuGeO${}_3$ is 
preferable to Zn-doped CuGeO${}_3$ 
for studies of impurity effects 
because of the more accurate control of $x$.
The use of Mg$^{2+}$ as a dilutant made it possible to observe 
unambiguously the phase transition between the D-AF and U-AF 
phases in impurity-doped CuGeO${}_3$.\cite{masuda} 
Careful treatment of Zn${}^{2+}$-doped samples also have 
made a similar observation possible in the present study. 

Measurements of the dc magnetic susceptibility were performed with a 
commercial 
SQUID magnetometer ($\chi$-MAG, Conductus Co., Ltd.). 
The synchrotron x-ray diffraction measurements were 
carried out at the MIT-IBM beamline 
X20A at the National Synchrotron Light Source (NSLS), Brookhaven National 
Laboratory.~\cite{yujie} 

\end{multicols}
\widetext
\begin{multicols}{2}
\narrowtext


\section{Experimental Results}

\begin{figure}[t]
\begin{center}
\includegraphics*[width=7.5cm]{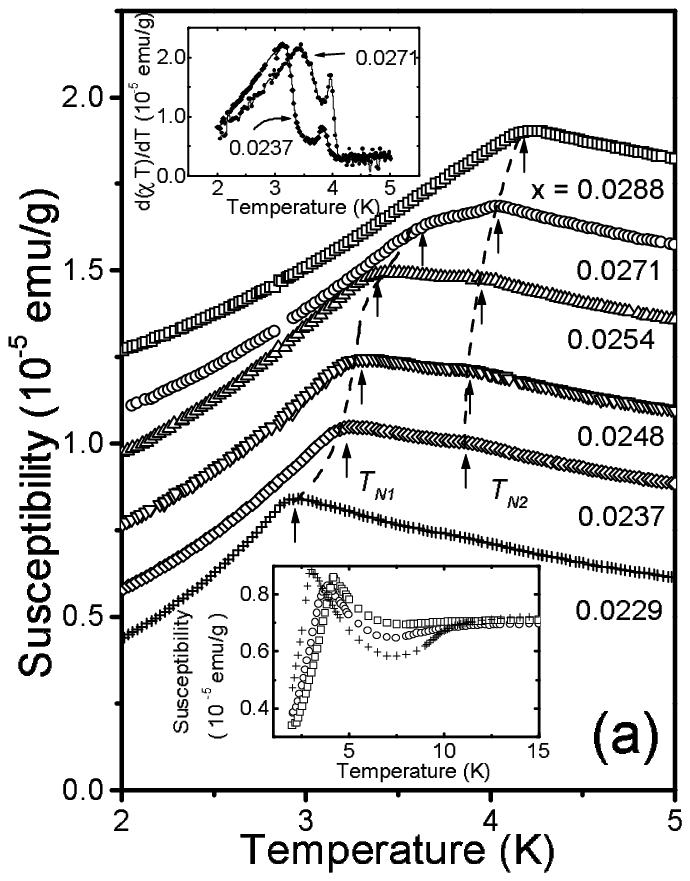}
\includegraphics*[width=7.5cm]{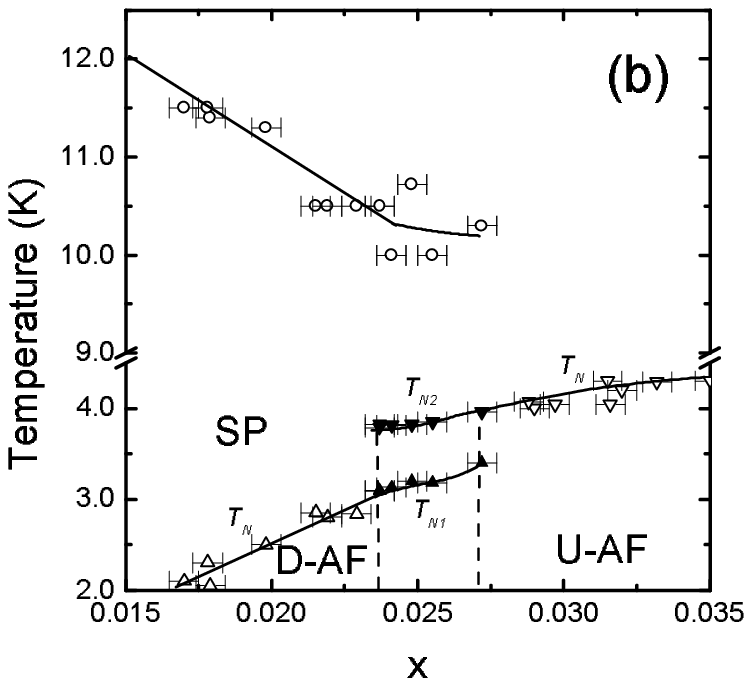}
\includegraphics*[width=7.5cm]{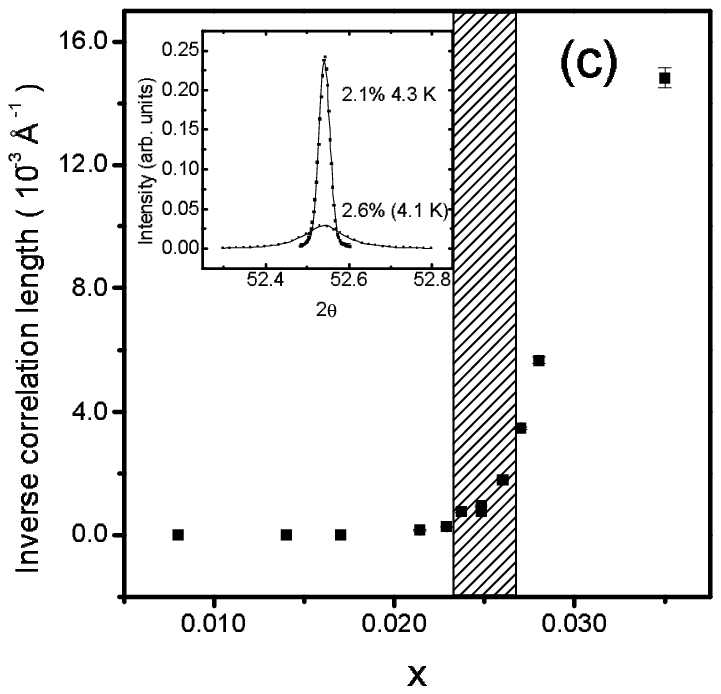}
\caption{
(a) The magnetic susceptibility of 
Cu${}_{1-x}$Mg${}_x$GeO${}_3$ near the AF transition temperature(s). 
The applied field is 1000 Oe. 
The data for different $x$ are shifted vertically. 
Double peaks are observed in the region of  
$0.0237 {<\atop\sim} x {<\atop\sim} 0.0271$, 
while a single peak is 
observed in the regions of  $x < 0.023$ and $0.028 < x$. 
The arrows indicate the anomaly due to AF transition. 
The inset in the upper left is $\partial(\chi_{\parallel}T)/\partial T$ 
for determining $T_{N1}$ and $T_{N2}$. 
The inset in the bottom shows the magnetic susceptibility in the same samples 
of $x$ = 0.0271, 0.0288, and 0.0299 for
2 K $< T <$ 20 K. 
(b) The temperature vs concentration phase diagram determined by 
the magnetic susceptibility measurements. 
Open circles, triangles, closed upward and downward 
triangles are $T_{SP}$, $T_N$, $T_{N1}$, and $T_{N2}$, respectively. 
Solid lines are guides to the eye. 
(c) Mg concentration dependence of the inverse correlation length of 
the lattice dimerization at $T =$ 4 K. The inset 
shows representative superlattice 
peak profiles. 
}
\label{fig1}
\end{center}
\end{figure}

The magnetic susceptibility in an applied field parallel 
to the $c$ axis 
($\chi_c(T)$) in Cu${}_{1-x}$Mg${}_x$GeO${}_3$ samples
around $x_c$ in the region of  2 K $< T <$ 5 K is shown in 
Fig.~\ref{fig1}(a). 
We observe clear double peaks in samples whose $x$'s are 
0.0237, 0.0248, 0.0254, 
and 0.0271, 
while only one sharp peak is observed in 0.0229 and 0.0288 samples. 
In contrast to the previous measurements,\cite{masuda} 
we took the data using temperature steps of 0.025 K; 
this reveals the detailed behavior of the susceptibility around 
$x \sim x_c$. 
The inset in the bottom of Fig~\ref{fig1}(a) 
shows $\chi_c(T)$ over a wider temperature range.
One can see the disappearance of the cusp 
in $\chi_c(T)$ around 10 K in the $x =$ 0.0288 sample while the cusp 
exists in the $x =$ 0.0271 
sample. 
This suggests that the SP transition still exists in the $x=0.0271$ sample 
and disappears in the $x=0.0288$ sample. 
Here we define $x_{c1}$ as the concentration $x$ where the double peaks 
first begin to 
appear, and $x_{c2}$ as the concentration
$x$, where only one peak begins to be observed
and at the same time the cusp in $\chi_c(T)$ corresponding to 
$T_{SP}$ disappears, with increasing $x$. 
With these definitions, we obtain $x_{c1} = 0.0237$ and $x_{c2} = 0.0271$. 
In cases when we do not distinguish $x_{c1}$ and $x_{c2}$, 
we use simply $x_c$, hereafter.
We determined $T_{SP}$ from the crossing points of linear functions fitted to 
$\chi_c(T)$ above and below the putative transition. 
We adopted Fisher's theory~\cite{fisher} to determine the AF 
transition temperature, according to which  
the AF transition temperature is signalled by a maximum in
$\partial(\chi_{\parallel}T)/\partial T$. 
When there are two peaks in $\partial(\chi_{\parallel}T)/\partial T$,
we define $T_{N1}$ and $T_{N2}$ as 
the maxima at lower and higher temperatures, respectively
(see the inset in the upper left of Fig.~\ref{fig1}(a)).       
If there is only one peak, we define the N\'eel temperature simply as $T_N$.
In this way we obtained the $T$ - $x$ phase diagram 
of Cu${}_{1-x}$Mg${}_x$GeO${}_3$ near $x_c$ 
shown in Fig.~\ref{fig1}(b). 

In Ref. ~\onlinecite{masuda} the existence of a 
first-order phase transition between the D-AF and U-AF phases 
was inferred from the 
observation of a sudden increase of $T_N$ at $x$ = 0.023, the broadening of 
$\chi_c(T)$ around $T_N$, and the disappearance of 
the cusp due to the SP transition. 
Instead of a single broad peak, we now observe clear double peaks.
$T_N$ below $x_{c1}$ is smoothly connected to $T_{N1}$ at $x_{c1}$,
while $T_{N2}$ is smoothly connected to $T_N$ ($x>x_{c2}$) at $x_{c2}$.  
Therefore $T_{N1}$ and $T_{N2}$ may be confidently assigned as the 
AF transition temperatures
with respect to the
D-AF and U-AF phases, respectively.
The lower $T_N$ curve never joins with the higher $T_N$ curve, which shows 
more directly the presence of 
the proposed first-order phase transition. 
Note that the double peaks are observed 
in the finite concentration region, $x_{c1} < x < x_{c2}$, 
where the lower and upper boundaries are separated
in $x$ by amounts well above 
our resolution of concentration. 

The phase transition between the D-AF and U-AF phases was 
also 
verified in 
neutron diffraction experiments\cite{nakao} but there remains some ambiguity 
in $x_c$. 
According to the neutron diffraction studies, $x_c$ was determined 
as approximately 0.027, 
which coincides with $x_{c2}$ of the 
present paper.
On the other hand, 
from the magnetic susceptibility measurements $x_c$ was deduced to be about 
0.023.\cite{masuda}
To determine at what temperature true SP-LRO is attained, synchrotron 
x-ray diffraction is a superior technique to neutron diffraction, 
because of its naturally very high resolution ($\sim$0.0002 $\rm\AA^{-1}$). 
The peak profiles of longitudinal scans at (1.5, 1, 1.5) of
samples with $x$ = 0.021 and 0.026 are 
shown in the inset of Fig.~\ref{fig1}(c). 
We observe superlattice peak in the samples with
$x \geq 0.023$
but the peak width is far wider than the resolution limit 
(5000 \AA${}$) even 
at low temperatures, 
that is, we determine that only SP-SRO exists in these samples. 
The concentration dependence of the inverse correlation length 
at 4 K is shown in Fig.~\ref{fig1}(c). 
The hatched zone corresponds to the double-peak region of the 
D-AF and U-AF phases 
determined by the 
magnetic susceptibility measurements. 
The correlation length ($\xi$) 
is larger than the resolution limit of 5000 $\rm\AA$\ at 
low temperatures for $x < x_{c1}$, 
decreases 
drastically 
in the double peak region,
and becomes much shorter at $x > x_{c2}$. 

\begin{figure}
\begin{center}
\includegraphics*[width=8cm]{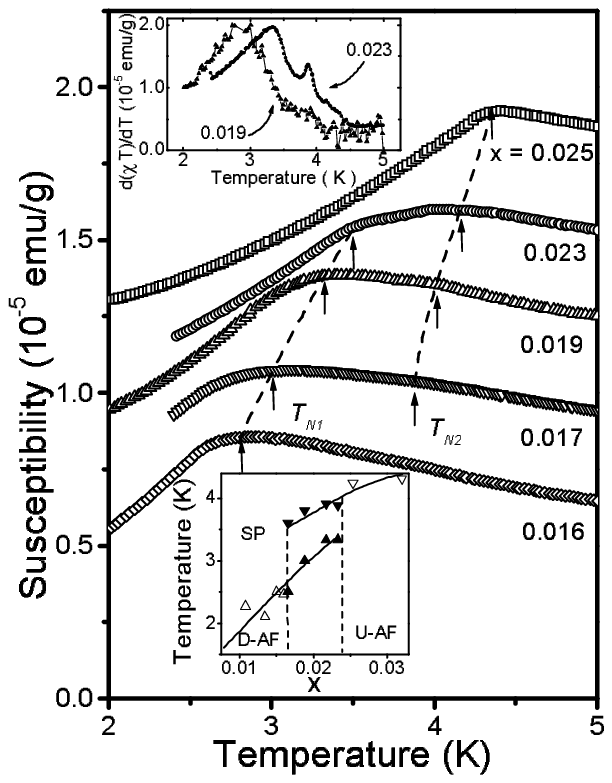}
\caption{
The magnetic susceptibility of Cu${}_{1-x}$Zn${}_x$GeO${}_3$ around $T_N$. 
Double peaks are observed 
in  $x = 0.019,$ and $0.023$ 
samples, 
while single 
peaks are observed 
in $x = 0.016$ and $0.025$
samples. 
In $x = 0.017$ the anomaly at $T_{N2}$ is confirmed 
by Fisher's method though it is not so clear in the 
raw susceptibility data. 
The inset in the bottom shows $T$ - $x$ phase diagram. 
The inset in the upper left is $\partial(\chi_{\parallel}T)/\partial T$. 
}
\label{fig2}
\end{center}
\end{figure}

We now discuss the Zn-doped CuGeO$_3$ system. As one might expect,
the phase transition between the D-AF and U-AF phases exists not only 
in the case of Mg-doped CuGeO${}_3$ but also in 
Zn-doped CuGeO${}_3$. 
$\chi_c(T)$ of Cu${}_{1-x}$Zn${}_x$GeO${}_3$ ($x$ = 0.016, 0.017, 
0.019, 0.023, and 
0.025) 
in the region of 
2 K $< T <$ 5 K is shown in Fig.~\ref{fig2}. 
Again we observe double peaks in samples with Zn$^{2+}$ concentrations of 
$x = 0.017$, 0.019 
and 0.023, while a single peak is observed in the 
$x$ = 0.016 and 0.025 samples; generally, the peak structure is not as 
clear as that of Cu${}_{1-x}$Mg${}_x$GeO${}_3$. 
$\partial(\chi_{\parallel}T)/\partial T$ is shown in the inset in upper 
left. 
The ambiguity seems to be due to the worse 
dilutant homogeneity in the Zn-doped samples compared with 
that of the Mg-doped samples 
as we explained in the previous section. 
In the case of Zn-doped CuGeO${}_3$ we 
obtain 
$x_{c1} \simeq 0.017$ and $x_{c2} \simeq 0.023$. 
The $T$ - $x$ phase diagram near $x_c$ is shown in the inset in the bottom. 
We observe a jump of $T_N$ in Zn-doped CuGeO$_3$ which is closely analogous 
to that in Mg-doped CuGeO${}_3$. 

\begin{figure}
\begin{center}
\includegraphics*[width=8cm]{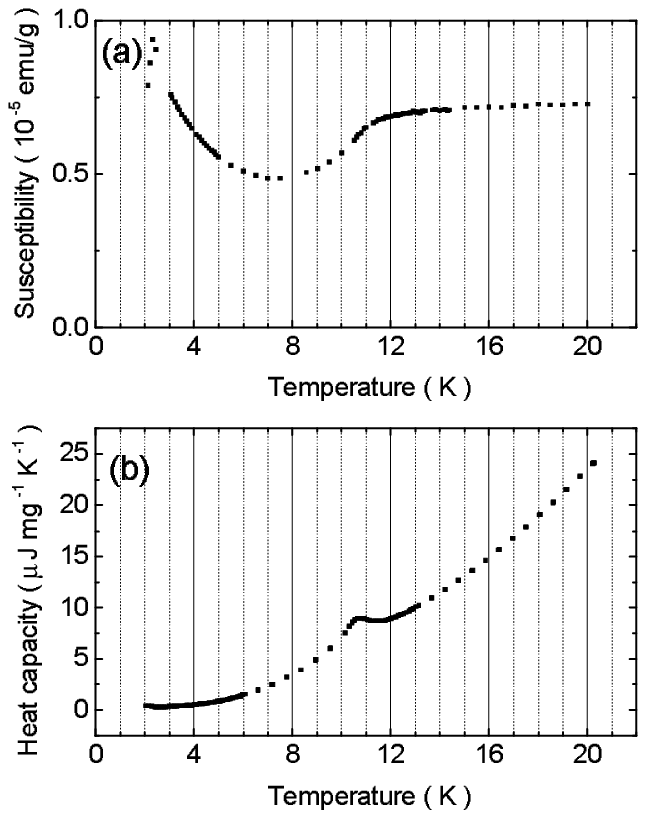}
\caption{
The experimental results of (a) the magnetic susceptibility 
and (b) heat capacity (b) in  Cu${}_{1-x}$Mg${}_x$GeO${}_3$ 
($x$ = 0.017) sample. 
}
\label{fig3}
\end{center}
\end{figure}

Now let us return to the SP transition in the Mg-doped CuGeO${}_3$ system. 
Figures~\ref{fig3}(a) and (b) show the experimental results 
of the magnetic susceptibility measurement and the heat-capacity 
measurement of 
Cu${}_{1-x}$Mg${}_x$GeO${}_3$ ($x$ = 0.017). 
We observe a cusp due to the SP transition at $T \sim 11.5$ K in 
the magnetic susceptibility (Fig.~\ref{fig3}(a)). 
As for heat capacity, we observe an 
anomaly 
due to the SP transition 
at $T \sim 10.7$ K (Fig.~\ref{fig3}(b)). 
From neutron diffraction measurements on the same concentration 
sample, it is found that the superlattice peak 
intensity begins to increase at $T \sim 10.8$ K in Fig. 2 of 
Ref.~\onlinecite{nakao}. 
The coincidence of these temperatures is good enough, 
and thus we concluded 
that $T_{SP}$ of the sample is about 11 K. 

On the other hand, high resolution x-ray diffraction  
measurements lead to rather different conclusions.
We observe that the peak width of the superlattice reflection is not
resolution limited at the $T_{SP}$ determined above
but rather it continues to narrow 
as the temperature is lowered 
below $T_{SP}$ as shown in Fig.~2 of Ref.~\onlinecite{yujie}. 
Here, we define the temperature where the SP correlations become long-range as 
$T_{SP}'$ and it is much lower than $T_{SP}$, especially for $x$ 
near $x_c$. 
The length 5000 \AA, corresponding to the resolution limit of the 
measurement, is much longer than the average impurity distance 
and consequently we may safely call 
the region of $T {<\atop\sim} T_{SP}'$ as the SP long-range order (LRO) region.

Since $T_{SP}'$ is lower than $T_{SP}$,  
true SP-LRO only exists at a temperature which is 
rather lower than the $T_{SP}$
deduced from the susceptibility measurements. 
The difference between $T_{SP}$ and $T_{SP}'$ does not reflect any
experimental artifact because the behavior of the x-ray integrated 
intensity shows almost the same temperature dependence as 
the peak intensity of the neutron diffraction. 
We should note that the resolution of neutron diffraction is 
much less ($\sim$200-500 \AA)\cite{nakao} and 
actually the peak intensity of 
the neutron diffraction corresponds to the 
integrated intensity of the x-ray 
diffraction within the temperature region of interest. 

\begin{figure}
\begin{center}
\includegraphics*[width=7cm]{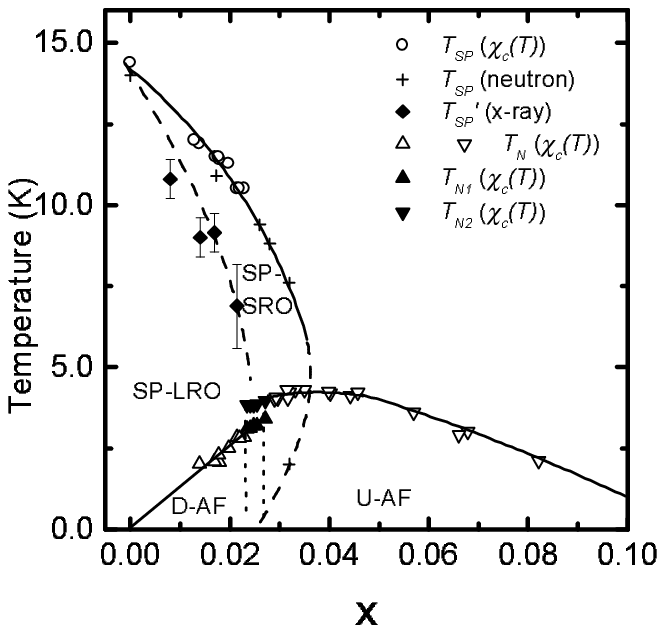}
\caption{
$T$ - $x$ phase diagram of Cu${}_{1-x}$Mg${}_x$GeO${}_3$ obtained 
by susceptibility measurements, x-ray diffraction and neutron 
diffraction.
Neutron diffraction data is from 
Fig.~6 
of Ref.~\protect\onlinecite{nakao}. 
}
\label{fig4}
\end{center}
\end{figure}

We show the $T$ - $x$ phase diagram of 
Cu${}_{1-x}$Mg${}_x$GeO${}_3$ in Fig.~\ref{fig4}. 
We have added $T_{SP}'$ as closed diamonds and $T_{SP}$ determined 
by the neutron diffraction~\cite{nakao} as 
plusses there. 
The $T_{SP}'$ decreases with $x$ 
and vanishes at $x_c \sim$ 0.024, i.e., the peak width does not tend to the 
resolution limit at low temperatures in the $x > x_c$ region. 

\section{Discussion}

We find clear double peaks in $\chi_c(T)$ 
(Fig.~\ref{fig1}(a), Fig.~\ref{fig2}) and a corresponding 
jump of $T_N$ in the $T$- $x$ phase diagram (Fig.~\ref{fig1}(b), 
the inset of Fig.~\ref{fig2}) 
of Cu${}_{1-x}M_x$GeO${}_3$ in the region $x_{c1} < x < x_{c2}$. 
While the jump of $T_N$ is the strongest evidence for the existence of 
a first-order phase transition between the D-AF and U-AF phases, 
the double peaks suggest the existence of spatial phase separation 
at a critical concentration, which is characteristic of a 
first-order phase transition. 
The measured phase separation region spreads over a finite region, 
$x_{c1} < x < x_{c2}$, 
which is consistent with the phase transition being of the 
first order. 
When a first-order phase transition occurs, metastable phenomena, 
e.g., 
supercooling and superheating,
appear around the critical point in general. 
The spread of the phase separation region over some non-zero range 
suggests the existence of metastable states as is indeed observed 
in the x-ray measurements.~\cite{yujie} 
We should note that this spread does not come from inhomogeneity 
of the impurity distribution in the case of Mg-doped CuGeO${}_3$, 
because the concentration fluctuation 
is within 0.1\% as shown in Fig.~\ref{fig0}. 

The fact that $T_{SP}'$ is much lower than $T_{SP}$ 
gives significant insight 
into 
how the SP order collapses with 
increasing impurity concentration.
$T_{SP}$ is the transition temperature inferred from
the dip in the magnetic susceptibility, the jump of the heat capacity, 
and the appearance of measurable diffraction intensity 
at the SP superlattice peak positions. 
The coincidence of SP transition temperatures, $T_{SP}$, in various
measurements, suggests 
that the correlation length of the dimerization of $\sim$500 \AA, 
which corresponds to the resolution limit of the neutron diffraction, 
is sufficient for the opening of the SP energy gap. 
In contrast to the case of pure CuGeO${}_3$ system where 
$T_{SP}$ and $T_{SP}'$ are the same,\cite{harris} 
the one-dimensional spin 
chain is cut at the impurity site in the Cu${}_{1-x}$$M_x$GeO${}_3$ system. 
For $T_{SP} > T > T_{SP}'$, the phase of dimerization is likely pinned
at the impurity sites, 
which is similar to the strong pinning interactions 
between the impurities and magnetic solitons which was suggested in 
the incommensurate phase 
in Cu${}_{1-x}$(Zn, Ni)${}_x$GeO${}_3$.\cite{valery} 
However, only local lattice rearrangements are needed to change the phase
at an impurity site. 
As the temperature decreases and the interchain interactions and 
spin-phonon coupling favoring the SP state become relatively more 
important, the individual finite SP domains begin to correlate with each
other over large distances, and at $T'_{SP}$ LRO is finally established. 
This model which could yield either a tricritical point with its concomitant 
first-order phase transition or more complicated reentrant scenario as in 
certain spin glass system is discussed briefly in Ref.~\onlinecite{yujie} 
and \onlinecite{valery2}. 

Therefore, for     
$T_{SP} > T > T_{SP}'$, the SP energy gap and the dimerization 
coexist though the lattice dimerization does not attain LRO.
Considering that the peak width determined in x-ray diffraction measurements
decreases gradually 
with decreasing temperature
and 
that 
there is no anomaly in the susceptibility, 
heat capacity, 
and neutron peak intensity at $T_{SP}'$, we cannot 
determine definitively whether 
the change from SP-SRO to SP-LRO is a true phase transition or a cross-over. 
The important 
point 
is that $T_{SP}'$, the temperature 
where 
$\xi$ is much longer 
than the average impurity distance, 
vanishes around $x_c$ (Fig.~\ref{fig1} (c) and Fig.~\ref{fig4}). 
The jump of $T_N$ at $x_{c1} < x < x_{c2}$ 
corresponds to the disappearance of 
SP-LRO, which is the evidence for the phase transition 
between the U-AF and D-AF phases by x-ray diffraction. 
When  $x$ is larger than $x_{c}$, 
SP-SRO is still present in the system. However, it should be considered as
the result of critical fluctuations of the true SP state found for
$x<x_c$.

Recently, Nakao 
{\it et al.}~\cite{nakao} have reported neutron diffraction measurements
performed on Mg-doped CuGeO$_3$ crystals. 
These group of authors, which included some of
the present authors, deduced the existence of the phase boundary 
between the 
D-AF and U-AF phases from 
the sudden change in SP lattice displacement $\delta$ 
and effective magnetic moment $\mu_{eff}$ at $x_c$. 
They have also proposed the existence of an intermediate phase with 
SP-SRO which is reentrant at low temperatures, and the existence of a
phase transition between the D-AF and U-AF phases at $T = 0$ K. 
This reentrancy has also been clearly seen in recent low temperature 
synchrotron x-ray measurements~\cite{valery2}. 

While some of the extant theories provide a qualitative description of
how impurity doping suppresses the SP phase and how the AF phase is 
induced~\cite{mostovoy,fabrizio}, 
they fail to explain the transition between the D-AF and U-AF
phases reported in our paper.
Saito\cite{saito} has recently proposed a model for the phase transition 
between the D-AF and U-AF phases 
at $T = 0$ K, and 
very recently she showed that the order of the phase transition 
depends on the ratio between the spin-lattice coupling and the 
interchain interaction.\cite{saito2} 
According to her work, the phase transition between the D-AF and U-AF phases 
can be of first order 
in the case of relatively large interchain interaction. 

While a detailed theoretical description of the two different AF phases
at non-zero temperatures is still absent, the behavior of $T_N$ as a 
function of $x$ can be qualitatively explained as follows.
The one-dimensionality of the 
spin interaction in CuGeO${}_3$ appears not to be as
good~\cite{nishi} 
as that of conventional organic SP materials.~\cite{bray} 
The $T$ - $x$ phase diagram 
of CuGeO${}_3$, therefore,  would be that of 
a typical diluted antiferromagnet, 
i.e, 
a monotonic decrease of $T_N$ with $x$ 
would 
be observed,
if the SP transition 
had not occurred 
in CuGeO${}_3$. 
Actually the occurrence of the SP phase suppresses the AF phase completely in 
pure CuGeO${}_3$. 
As $x$ increases, the SP phase is suppressed and 
the D-AF 
phase develops at an infinitesimally small impurity 
concentration,~\cite{manabe} and 
at $x_c$ the SP-LRO disappears and the phase transition 
from the D-AF to U-AF phase occurs. 
Since the SP-SRO, however, still exists above $x_c$, the AF phase is 
suppressed, 
and $T_N$ exhibits a plateau for $x_c < x {<\atop\sim} 0.04$. 
For $x {>\atop\sim} 0.04$ any SP-SRO is very weak (Ref.~\onlinecite{nakao}) and 
therefore typical behavior of a diluted antiferromagnet, 
that is, a
monotonic 
decrease of $T_N$ with $x$, is observed.


\section{Concluding remarks}

We have confirmed the phase transition between the U-AF and D-AF phases 
in Mg-doped CuGeO$_3$ by detailed susceptibility measurements 
and high resolution synchrotron x-ray diffraction studies. 
The results of previous neutron diffraction experiments
also suggested a similar phase transition. 
We found clear double peaks in the magnetic susceptibility around 
$x \sim x_c$. We have interpreted these peaks as the result of two 
separate N\'eel transitions, in which case 
spatial phase separation between the D-AF and U-AF phases is present in the
system. These features are interpreted as the result of
an intrinsic first order transition. 

Our x-ray diffraction measurements show that the SP dimerization attains
long-range order only for $x<x_c$. Thus, the transition from the D-AF
to U-AF phase is characterized by the loss of the SP long-range order.

Our susceptibility measurements show that
Cu${}_{1-x}$Zn${}_x$GeO${}_3$ exhibits the same kind of behavior 
as Cu${}_{1-x}$Mg${}_x$GeO${}_3$, 
and the $T$ - $x$ phase diagrams of these compounds are very similar.
We, therefore, conclude that the D-AF and U-AF phases, and the corresponding
transition between these phases at $x_c$ are also present in Zn-doped
CuGeO$_3$. 

\section{acknowledgements}

We would like to thank H. Nakao, Y. Fujii, 
M. Nishi, 
K. Hirota, 
and G. Shirane for helpful discussions of the neutron diffraction experiments. 
We are also grateful to M. Saito for valuable discussion and 
for providing a preprint prior to publication. 
We would like to thank H. Fukuyama for general discussions on the 
role of disorder in spin-gap systems. 
We also acknowledge A. Fujioka for the growth of some of the 
samples used. 
We thank S. LaMarra for assistance with the synchrotron experiments.
This work is supported in part by a 
Grant-in-Aid for COE Research ``SCP coupled system" from the Ministry of 
Education, Science, Sports and Culture of Japan
and by a Research Fellowship of the Japan Society 
for the Promotion of Science for Young Scientist (T. M.). 
This work was also supported by the NSF under Grant No. DMR97-04532.


\end{multicols}

\end{document}